\documentclass{article}
\usepackage[english]{babel}
\usepackage[letterpaper,top=2cm,bottom=2cm,left=3cm,right=3cm,marginparwidth=1.75cm]{geometry}
\usepackage{amsmath}
\usepackage{graphicx}

\usepackage[natbibapa]{apacite} 
\usepackage[colorlinks=true, allcolors=OliveGreen]{hyperref} 
\usepackage[svgnames,usenames,dvipsnames]{xcolor} 
\usepackage{setspace} 
\usepackage{threeparttable} 
\usepackage{authblk} 
\usepackage{indentfirst} 
\usepackage{titlesec} 

\title{Teacher Mental Health During the COVID-19 Pandemic: Informing Policies to Support Teacher Well-being and Effective Teaching Practices}

\author{\vspace{10mm}Joseph M. Kush\textsuperscript{1}\footnote{Correspondence concerning this article should be addressed to Joseph Kush, Johns Hopkins Bloomberg School of Public Health, 624 N Broadway Room 841, Baltimore, MD 21205. Email: jkush1@jhu.edu}, Elena Badillo-Goicoechea\textsuperscript{1}, Rashelle J. Musci\textsuperscript{1}, and Elizabeth A. Stuart\textsuperscript{1}}

\affil{\vspace{10mm}\textsuperscript{1}Department of Mental Health, Johns Hopkins Bloomberg School of Public Health}

\date{\vspace{10mm}September 3, 2021}

\begin{document}
\setstretch{1.25}
\maketitle

\begin{abstract}
\noindent
While there is an emergence of research investigating the educational impacts of the COVID-19 pandemic, empirical studies assessing teacher mental health throughout the pandemic have been scarce. Using a large national dataset, the current study first compared mental health outcomes during the pandemic between pK-12 teachers and professionals in other occupations. Further, we compared the prevalence of mental health outcomes between in-person and remote teachers (n = 131,154). Findings indicated teachers reported greater mental health concerns than those in other professions, and that remote teachers reported significantly higher levels of distress than those teaching in-person. Policy implications are discussed, with a focus on providing support to meet the evolving needs of teachers. 
\end{abstract}

\section{Introduction}
\par
School districts across the United States faced an unprecedented disruption during the spring of 2020 and the 2020/2021 academic year due to the COVID-19 pandemic. Although the initial response of most districts was to switch to fully virtual learning (\citeauthor{national2020reopening}, \citeyear{national2020reopening}), schools varied in their instructional policies throughout the course of the pandemic: pivoting between fully remote, full onsite, and a combination of remote and onsite, sometimes allowing both students and teachers in school buildings, and other times not. While there is an ongoing emergence of research investigating the effects of the COVID-19 pandemic on student achievement (e.g., \shortcites{bailey2021achievement}\citeauthor{bailey2021achievement}, \citeyear{bailey2021achievement}; \citeauthor{kuhfeld2020projecting}, \citeyear{kuhfeld2020projecting}), student mental health (e.g., \citeauthor{prowse2021coping}, \citeyear{prowse2021coping}; \shortcites{tortella2021mindfulness}\citeauthor{tortella2021mindfulness}, \citeyear{tortella2021mindfulness}; \shortcites{wasil2021promoting}\citeauthor{wasil2021promoting}, \citeyear{wasil2021promoting}), and public health and safety measures in academic settings (e.g., \citeauthor{lessler2021household}, \citeyear{lessler2021household}; \citeauthor{doyle2021covid}, \citeyear{doyle2021covid}; \shortcites{ismail2021sars}\citeauthor{ismail2021sars}, \citeyear{ismail2021sars}), there has been less focus on teacher mental health during the pandemic and how the instructional modalities, and changes in them, might relate to teacher mental health.

\par
Even before the challenges brought about by the pandemic, teacher stress has long been a major concern, with teachers consistently experiencing some of the highest levels of occupational stress among most occupations (\citeauthor{johnson2005experience}, \citeyear{johnson2005experience}; \shortcites{markow2013metlife}\citeauthor{markow2013metlife}, \citeyear{markow2013metlife}).With the quick initial pivot to remote teaching, followed by uncertainty due to modifications of instructional policies, we might expect that there would be particularly high levels of stress and negative mental health outcomes among teachers during the pandemic. As districts continue to grapple with decisions regarding in-person and remote modalities as the ongoing public health emergency evolves, teachers face even higher uncertainty. More research is needed to understand teacher mental health during the pandemic, in an effort to identify groups in need of additional supports and potential interventions in the context of necessary public health containment measures, such as school closures.

\par
The current study aims to address this gap in the literature by examining associations between the COVID-19 pandemic and teachers’ mental health among primary and secondary school teachers across the United States. Utilizing a large national dataset, we first examined differences in mental health during the pandemic between pre-Kindergarten through 12th grade (pK-12) teachers and other professionals, including healthcare workers, office workers, and “other” professionals. Second, we compared differences in mental health outcomes between individuals teaching in-person and remote modalities. Findings from our study contribute to the growing body of literature on the educational impacts of the pandemic by focusing explicitly on teachers and teacher mental health, and is the first study to do so using a large national dataset. 

\section{Teacher Stress}
\par
Teachers experience some of the highest levels of occupational stress and lowest levels of well-being among all professions (\citeauthor{johnson2005experience}, \citeyear{johnson2005experience}; \citeauthor{Bauer2006correlation}, \citeyear{Bauer2006correlation}). Nearly one in four public school teachers in elementary and secondary schools in the U.S. indicates stress as a reason to discontinue teaching, a number that continues to increase over time (\shortcites{snyder2019digest}\citeauthor{snyder2019digest}, \citeyear{snyder2019digest}). \shortcites{markow2013metlife}\cite{markow2013metlife} found that 51\% of teachers U.S. K-12 public school teachers report “feeling under great stress several days a week”, culminating in the lowest levels of teacher satisfaction in 25 years. Greenberg and colleagues (\citeyear{greenberg2016teacher}) identified four main sources of stress for teachers, including school organization (e.g., school culture), job demands (e.g., high-stakes testing), work resources (e.g., autonomy and decision-making), and personal resources (e.g., social-emotional competence). Unfortunately, as a result of these often-overwhelming burdens, nearly half of all teachers end up leaving the profession within their first five years (\shortcites{pas2012teacher}\citeauthor{pas2012teacher}, \citeyear{pas2012teacher}). 

\par
Beyond high turnover rates, the negative consequences of teachers’ stress are wide-ranging and may contribute to negative mental health outcomes such as feelings of burnout, depression and anxiety, and poor coping styles (\shortcites{herman2018empirically}\citeauthor{herman2018empirically}, \citeyear{herman2018empirically}; \shortcites{maslach2001job}\citeauthor{maslach2001job}, \citeyear{maslach2001job}). Moreover, teachers’ feelings of occupational stress ultimately impact their ability to employ effective teaching practices, resulting in negative impacts on student behavior (\shortcites{hoglund2015classroom}\citeauthor{hoglund2015classroom}, \citeyear{hoglund2015classroom}; \citeauthor{skaalvik2007dimensions}, \citeyear{skaalvik2007dimensions}) and student achievement (\shortcites{braun2019middle}\citeauthor{braun2019middle}, \citeyear{braun2019middle}; \shortcites{tsouloupas2010exploring}\citeauthor{tsouloupas2010exploring}, \citeyear{tsouloupas2010exploring}). Recent work by Braun and colleagues (\citeyear{braun2019middle}; \citeyear{braun2020effects}) has examined the role of teachers’ emotion regulation skills in the workplace and found that teachers who used cognitive reappraisal to regulate their emotions had students with more prosocial behavior. \cite{mclean2015depressive} and \shortcites{roberts2016exploring}\cite{roberts2016exploring} found that depressive symptomology among teachers was related to poorer student achievement and social-emotional development. \shortcites{ortega2021psychological}\cite{ortega2021psychological} demonstrated that psychological flexibility, or the ability to experience the present moment with an open mind, mediated the relationship between loneliness and stress among professors.

\subsection{Depression}
\par
Depressive symptoms have been found to be related to the quality of teachers’ interactions with students \citep{pianta2005features}. Depression causes persistent feelings of sadness, leading to emotional and physical problems such as loss of interest in activities, loss of energy, or a decreased ability to function at work \citep{american2013highlights}. Prior research has demonstrated that teachers with higher levels of depression had fewer effective interactions with students \citep{hamre2004self}, and that students in classrooms with more depressed teachers had greater rates of problem behaviors and fewer social skills \citep{roberts2016exploring}.  Other work has also found that teachers’ depressive symptoms were related to poorer student math achievement \citep{mclean2015depressive}.

\subsection{Anxiety}
\par
Considerably less research has focused on facets of teacher mental health other than depression, such as anxiety. Further, much of the prior work examining anxiety in teachers has dealt with teaching anxiety specifically, rather than a generalized symptomatology. Anxiety disorders involve a persistent and excessive fear or worry and are often accompanied by symptoms such as restlessness, muscle tension, or difficulty focusing \citep{pianta2005features}. Early research by \cite{sinclair1987teacher} found that anxiety in teachers often leads to increased stress in their students and poorer student evaluations. More recent work has found that reduced anxiety in teachers is related to improved teaching efficacy beliefs and commitment (\citeauthor{al2011path}, \citeyear{al2011path}; \shortcites{daniels2011effect}\citeauthor{daniels2011effect}, \citeyear{daniels2011effect}). \cite{brackett} found that anxiety was by far the most frequent emotion mentioned by teachers in surveys during the COVID-19 crisis, and suggest that teachers who express anxiety are more likely to alienate students, ultimately impacting students’ sense of safety.

\subsection{Isolation}
\par
Many researchers have noted that social isolation remains a vital, yet underrepresented area of research and importance (\citeauthor{seeman2000health}, \citeyear{seeman2000health}; \citeauthor{smith2020covid}, \citeyear{smith2020covid}). This may be especially true given the COVID-19 pandemic, in which actions such as physical distancing, isolation, and quarantine help prevent the spread of the virus \citep{centers2021guidance}. The negative effects of social isolation on overall health and well-being have been shown to be consistently associated with premature mortality \citep{hawkley2010loneliness}, depression \shortcites{shankar2011loneliness}\citep{shankar2011loneliness}, and substance use \shortcites{holt2015loneliness}\citep{holt2015loneliness}. Specific to education, researchers have documented feelings of isolation among students in distance learning environments (\shortcites{croft2010overcoming}\citeauthor{croft2010overcoming}, \citeyear{croft2010overcoming}; \citeauthor{liu2008student}, \citeyear{liu2008student}) and evidence suggests professional isolation among teachers is related to worse relationships with colleagues \shortcites{dussault1999professional}\citep{dussault1999professional}. 

\section{Policy Decisions During the COVID-19 Pandemic}
\par
When the pandemic began in the spring of 2020, schools were operating with limited guidance given the unprecedented situation, resulting in an initial widespread closure of in-person instruction (\shortcites{honein2021data}\citeauthor{honein2021data}, \citeyear{honein2021data}). Decisions regarding how to educate safely and effectively varied at the district, county, or state level, often with the added option of allowing parents to choose between onsite or remote instruction \citep{lupton2021consistency}. In spite of their crucial role in helping to contain the spread of COVID-19, school closures pose some risks and challenges. These decisions are extremely multifaceted, in which leaders must weigh and consider public health and safety, the feasibility of mitigation measures, and the transition to and implementation of effective remote instruction, among other considerations \citep{national2020reopening}. Moreover, attention to heterogenous and marginalized student populations must also be integrated into such decisions. As highlighted by
\cite{armitage2020considering}, school closures have the potential to exacerbate pre-existing inequalities, often disproportionately affecting disadvantaged children. Such examples may include students from homes with food insecurity who rely on school for meals, or students from homes without access to a computer or wireless internet 
\citep{van2020covid}. It is crucial to understand how these decisions have affected student learning, public health and safety, as well as student and teacher mental health. Overall, more research is needed to understand the impacts of these policy decisions on teachers’ mental health during the pandemic. 

\section{Current Study}
\par
The aim of the current study was to deepen our understanding of the associations between the COVID-19 pandemic and teachers’ mental health. First, we examined differences in mental health during the pandemic between teachers and professionals in other occupations. Building upon prior research (\citeauthor{johnson2005experience}, \citeyear{johnson2005experience}; \citeauthor{kahn1993caring}, \citeyear{kahn1993caring}), we specifically compared the stress of pK-12 teachers to professionals in healthcare, office, and “other” type occupations. Second, we compared mental health during the pandemic between teachers teaching in in-person vs. remote modalities. Following findings by \cite{johnson2005experience}, we hypothesized that teachers would present higher distress, compared to other professions. We further hypothesized that teachers teaching in person would show lower distress than those teaching virtually/remotely, as social interactions at school have been shown to contribute to teacher satisfaction 
(\shortcites{veldman2016veteran}\citeauthor{veldman2016veteran}, \citeyear{veldman2016veteran}). 

\section{Methods}
\subsection{Study Design and Procedure}
\par
Data came from the COVID-19 Symptom Survey, a large online survey developed in collaboration between Carnegie Mellon University’s Delphi Group and Facebook (\citeauthor{delphi}, \citeyear{delphi}; \citeauthor{salomon2021us}, \citeyear{salomon2021us}). This cross-sectional daily survey invites a stratified random sample of Facebook users to take the survey, and consists of questions related to health symptoms, preventive behaviors, mental health, and more
\citep{salomon2021us}. We used data from adult (18 years or older) participants who responded to the survey from September 8th, 2020 until March 28th, 2021 (\textit{N} = 5,186,378). Table 1 provides demographic information by job type across all employed respondents (\textit{N} = 2,775,974) and by in-person and remote modality for teachers only (\textit{N} = 135,488).

%
\begin{table}[!ht]
\centering
\small
\resizebox{\textwidth}{!}{\begin{tabular}{llllllllll}
\multicolumn{10}{l}{Table 1}   \\
\multicolumn{10}{l}{Sociodemographic   Factors for All Professionals (September 2020 – March 2021)}                                         \\
\hline \\
\multicolumn{2}{l}{}              & \multicolumn{5}{l}{All Professions (\%)}                  & \multicolumn{3}{l}{pK-12 Teachers (\%)} \\
\cline{3-10} \\
\multicolumn{2}{l}{Variable}      & Education & Healthcare & Office & Other & \textit{p}              & In-person   & Remote   & \textit{p}               \\
\hline \\
\multicolumn{2}{l}{Sample Size}      & 135, 488 & 427,117 & 351,632 & 1,861,737 &                & 106,000 & 28,693        &                \\
\multicolumn{2}{l}{Gender}           &          &         &         &           &                &         &               &                \\
 & Female                            & 74.0     & 72.0    & 65.0    & 48.0      &                & 74.0    & 74.0          &                \\
 & Male                              & 25.0     & 26.0    & 34.0    & 50.0      &                & 25.0    & 25.0          &                \\
 & Non-binary                        & 0.5      & 0.6     & 0.6     & 0.9       &                & 0.5     & 0.7           &                \\
 & Other/prefer not to   answer   & 0.8       & 1.4        & 0.9    & 1.9    & \textless .001 & 0.8         & 0.8      & \textless .001  \\
\multicolumn{2}{l}{Age}              &          &         &         &           &                &         &               &                \\
 & 18-24                             & 8.2      & 9.4     & 7.1     & 12.0      &                & 8.9     & 5.6           &                \\
 & 25-34                             & 21.0     & 21.0    & 19.0    & 20.0      &                & 21.0    & 21.0          &                \\
 & 35-44                             & 23.0     & 21.0    & 22.0    & 21.0      &                & 23.0    & 25.0          &                \\
 & 45-54                             & 26.0     & 22.0    & 24.0    & 18.0      &                & 26.0    & 27.0          &                \\
 & 55-64                             & 16.0     & 19.0    & 20.0    & 18.0      &                & 15.0    & 17.0          &                \\
 & 65+                               & 5.1      & 7.4     & 7.9     & 23.0      & \textless .001 & 5.0     & 5.1           & \textless .001 \\
\multicolumn{2}{l}{Education Level}  &          &         &         &           &                &         &               &                \\
 & Less than HS                      & 0.1      & 0.8     & 0.3     & 2.3       &                & 0.1     & \textless 0.1 &                \\
 & HS                                & 1.8      & 8.7     & 8.9     & 15.0      &                & 1.9     & 1.3           &                \\
 & Some college                      & 5.8      & 22.0    & 26.0    & 25.0      &                & 6.2     & 4.1           &                \\
 & College/professional   degree     & 66.0     & 58.0    & 57.0    & 48.0      &                & 66.0    & 64.0          &                \\
 & Graduate degree                   & 26.0     & 11.0    & 7.0     & 9.2       & \textless .001 & 25.0    & 30.0          & \textless .001 \\
\multicolumn{2}{l}{Metro Size}       &          &         &         &           &                &         &               &                \\
 & Not adjacent to metro   area      & 4.9      & 5.0     & 3.9     & 4.7       &                & 5.7     & 2.5           &                \\
 & Adjacent to metro   area          & 9.1      & 9.9     & 7.6     & 9.3       &                & 11.0    & 5.1           &                \\
 & Fewer than 250,000                & 11.0     & 12.0    & 9.6     & 11.0      &                & 11.0    & 6.6           &                \\
 & 250,000 to 1 million   population & 27.0     & 28.0    & 27.0    & 27.0      &                & 28.0    & 26.0          &                \\
 & 1 million population   or more & 48.0      & 45.0       & 52.0   & 49.0   & \textless .001 & 44.0        & 62.0     & \textless .001  \\
\multicolumn{2}{l}{Mental Health}    &          &         &         &           &                &         &               &                \\
 & Depressive symptoms               & 12.0     & 13.0    & 12.0    & 14.0      &                & 12.0    & 13.0          &                \\
 & Anxiety symptoms                  & 24.0     & 19.0    & 19.0    & 17.0      &                & 24.0    & 23.0          &                \\
 & Feelings of Isolation             & 18.0     & 17.0    & 19.0    & 21.0      & \textless .001 & 17.0    & 23.0          & .05           
\\ 
\hline
\end{tabular}}

\begin{tablenotes}
  \small
  \item Note: Other professionals are defined as either management, business and financial operations, computer and mathematical, architecture and engineering, life, physical, and social science, legal, farming, fishing, and forestry, military, or any other occupational group. Teachers include only the elementary, middle, or secondary school teachers, among all respondents in the “Education, training, and library” workers category. Data source: U.S. COVID-19 Symptom Survey, \cite{delphi}; \cite{dong2020interactive}.
\end{tablenotes}
\end{table}
%

\subsection{Measures}
\par
Three measures of mental health were examined: 1) depressive symptoms, 2) anxiety symptoms, and 3) feelings of isolation. All three items shared the following question stem: “In the past 7 days, how often have you ...”. The three items were originally scored along a four-point Likert scale with responses ranging from 1 = none of the time to 4 = all of the time. We recoded each item into a dichotomous indicator, where 0 = none or some of the time, and 1 = most or all of the time. 
Regarding occupation, survey respondents were asked to indicate the occupational group that best fits the main kind of paid work they were engaged in the previous four weeks. This allowed us to compare mental health outcomes between teachers and other professional groups (Aim 1), and between teachers teaching in in-person vs. remote modalities (Aim 2). In both cases, we defined “teacher” as being a pre-Kindergarten, elementary, middle, or secondary teacher. We defined a comparison group consisting of healthcare (e.g., nurses, physicians, dentists), office (e.g., customer service representatives, administrative support), and “other” (e.g., military, legal, farming) workers. 
The distinction between in-person and remote modality was made by using the respondent’s answer (asked to all respondents who reported being paid for work during the last month) to the survey question of whether or not they had worked outside their home during that same period. Sociodemographic characteristics (e.g., gender, age, education level, number of children, household size, and level of financial worry) were also collected in the survey as included in the models as covariates. We also controlled for a set of county-level covariates: urbanicity (U.S. Census), COVID-19 cases and deaths (lagged by two weeks; JHU Coronavirus Resource Center, \shortcites{dong2020interactive}\citeauthor{dong2020interactive}, \citeyear{dong2020interactive}), in addition to including U.S. State and month as fixed effects.

\subsection{Analyses}
\par
We first assessed potential differences in mental health outcomes between teachers and healthcare, office, and other workers during the pandemic, in terms of their probability of reporting a negative mental health outcome.  These are examined using logistic regression of each outcome as a function of profession and the individual- and county-level covariates. We then compared outcomes between individuals teaching in in-person vs. remote modalities, again using logistic regressions, but fit only among teachers and with in-person status the key predictor of interest. Each model was weighted for non-response and coverage bias following the weighting scheme outlined in \cite{salomon2021us}. All models were estimated using the survey package in R \citep{lumley2004analysis}, with a survey design stratified by state. The final sample for each model included only complete cases for outcome, exposure, and all covariates. For Aim 1 analyses, this corresponded to 70.4\% of the sample. For Aim 2, this corresponded to 68.3\% of the sample.

\section{Results}
\subsection{Comparison of Outcomes Among Teachers and Other Professionals}
\par
Logistic regression results indicated healthcare workers (OR = 0.91, \textit{p} $<$ .001; OR = 0.70, \textit{p} $<$ .001), office workers (OR = 0.99, \textit{p} $<$ .001; OR = 0.82, \textit{p} $<$ .001), and other workers (OR = 0.96, \textit{p} $<$ .001; OR = 0.79, \textit{p} $<$ .001) were significantly less likely to report depression and anxiety symptoms than teachers, although the point estimates were sometimes very close to 1 (see Table 2). Similarly, healthcare workers (OR = 0.94, \textit{p} $<$ .001) were significantly less likely to report feeling isolated than teachers, although office workers and other workers were significantly more likely to report isolation (OR = 1.18, \textit{p} $<$ .001; OR = 1.08, \textit{p} $<$ .001). Figure 1 provides a graphical display of the probability of mental distress across the four occupation groups.

%
\begin{table}[!ht]
\centering
\small
\resizebox{\textwidth}{!}{\begin{tabular}{llllllll}
\multicolumn{8}{l}{Table 2}                                                                                                                                   \\
\multicolumn{8}{l}{Results for   Prevalence of Mental Health Outcomes (Logistic regression): Teachers vs.   Other Professionals}                              \\
\hline \\
\multicolumn{2}{l}{Variable} &
  \multicolumn{2}{l}{Depressive Symptoms} &
  \multicolumn{2}{l}{Anxiety Symptoms} &
  \multicolumn{2}{l}{Feelings of Isolation} \\
\hline \\
\multicolumn{2}{l}{}                                                      & OR    & CI (95\%)         & OR    & CI (95\%)         & OR    & CI (95\%)         \\
\cline{3-8} \\
\multicolumn{2}{l}{Profession (reference:   Teachers)}                    &       &                   &       &                   &       &                   \\
                 & Healthcare                                             & 0.91  & {[}0.89, 0.93{]}  & 0.70  & {[}0.69, 0.72{]}  & 0.94  & {[}0.91, 0.96{]}  \\
                 & Office                                                 & 0.99  & {[}0.97, 1.02{]}  & 0.82  & {[}0.80, 0.84{]}  & 1.18  & {[}1.15, 1.21{]}  \\
                 & Other                                                  & 0.96  & {[}0.94, 0.98{]}  & 0.79  & {[}0.77, 0.80{]}  & 1.08  & {[}1.06, 1.10{]}  \\
\multicolumn{2}{l}{Gender (reference:   Female)}                          &       &                   &       &                   &       &                   \\
                 & Male                                                   & 0.71  & {[}0.70, 0.72{]}  & 0.53  & {[}0.52, 0.53{]}  & 0.81  & {[}0.81, 0.82{]}  \\
 &
  Non-binary &
  3.17 &
  {[}3.01,   3.33{]} &
  2.46 &
  {[}2.34,   2.58{]} &
  2.32 &
  {[}2.21,   2.43{]} \\
 &
  Other/prefer not to   answer &
  1.25 &
  {[}1.19,   1.32{]} &
  0.87 &
  {[}0.83,   0.91{]} &
  1.16 &
  {[}1.11,   1.21{]} \\
\multicolumn{2}{l}{Age (reference: 65+)}                                  &       &                   &       &                   &       &                   \\
                 & 18-24                                                  & 7.14  & {[}6.88, 7.40{]}  & 7.58  & {[}7.34, 7.82{]}  & 4.43  & {[}4.30, 4.56{]}  \\
                 & 25-34                                                  & 4.44  & {[}4.30, 4.59{]}  & 5.23  & {[}5.09, 5.37{]}  & 3.08  & {[}3.00, 3.15{]}  \\
                 & 35-44                                                  & 2.82  & {[}2.73, 2.91{]}  & 3.49  & {[}3.39, 3.59{]}  & 2.31  & {[}2.26, 2.37{]}  \\
                 & 45-54                                                  & 1.92  & {[}1.86, 1.98{]}  & 2.32  & {[}2.25, 2.38{]}  & 1.67  & {[}1.63, 1.72{]}  \\
                 & 55-64                                                  & 1.42  & {[}1.37, 1.47{]}  & 1.62  & {[}1.57, 1.66{]}  & 1.36  & {[}1.33, 1.39{]}  \\
\multicolumn{2}{l}{Education Level (reference:   Less than HS)}           &       &                   &       &                   &       &                   \\
                 & HS                                                     & 0.95  & {[}0.90, 1.00{]}  & 0.97  & {[}0.93, 1.02{]}  & 0.90  & {[}0.86, 0.94{]}  \\
                 & Some college                                           & 1.01  & {[}0.96, 1.06{]}  & 1.14  & {[}1.10, 1.20{]}  & 1.10  & {[}1.05, 1.15{]}  \\
                 & College/professional   degree                          & 0.86  & {[}0.82, 0.90{]}  & 1.13  & {[}1.09, 1.18{]}  & 1.18  & {[}1.12, 1.23{]}  \\
                 & Graduate degree                                        & 0.83  & {[}0.79, 0.87{]}  & 1.15  & {[}1.10, 1.20{]}  & 1.34  & {[}1.28, 1.40{]}  \\
\multicolumn{2}{l}{Metro Size (reference:   Metro \textgreater 1 million} &       &                   &       &                   &       &                   \\
                 & Nonmetro-Not adjacent   to metro area                  & 0.93  & {[}0.90, 0.95{]}  & 0.92  & {[}0.90, 0.95{]}  & 0.82  & {[}0.80, 0.84{]}  \\
                 & Nonmetro-Adjacent to   metro area                      & 0.97  & {[}0.95, 0.99{]}  & 0.93  & {[}0.91, 0.95{]}  & 0.85  & {[}0.83, 0.86{]}  \\
                 & Metro-Fewer than   250,000                             & 1.03  & {[}1.01, 1.05{]}  & 0.98  & {[}0.97, 1.00{]}  & 0.91  & {[}0.90, 0.93{]}  \\
                 & Metro-250,000 to 1   million population                & 1.04  & {[}1.02, 1.06{]}  & 1.00  & {[}0.99, 1.01{]}  & 0.97  & {[}0.96, 0.98{]} 
\\
\hline
\end{tabular}}

\begin{tablenotes}
  \small
  \item Note: OR = odds ratio. Only estimated coefficients for exposure and sociodemographic covariates are show. Models also adjusted for number of children and elders in the household, financial worry, and the number of COVID-19 positive people respondents knew, urbanicity, COVID-19 cases and deaths (lagged by 2 weeks), and fixed-effects for U.S. State and month.
\end{tablenotes}

\end{table}
%

\begin{figure}[!ht]
  \centering
  \includegraphics[width=1\columnwidth]{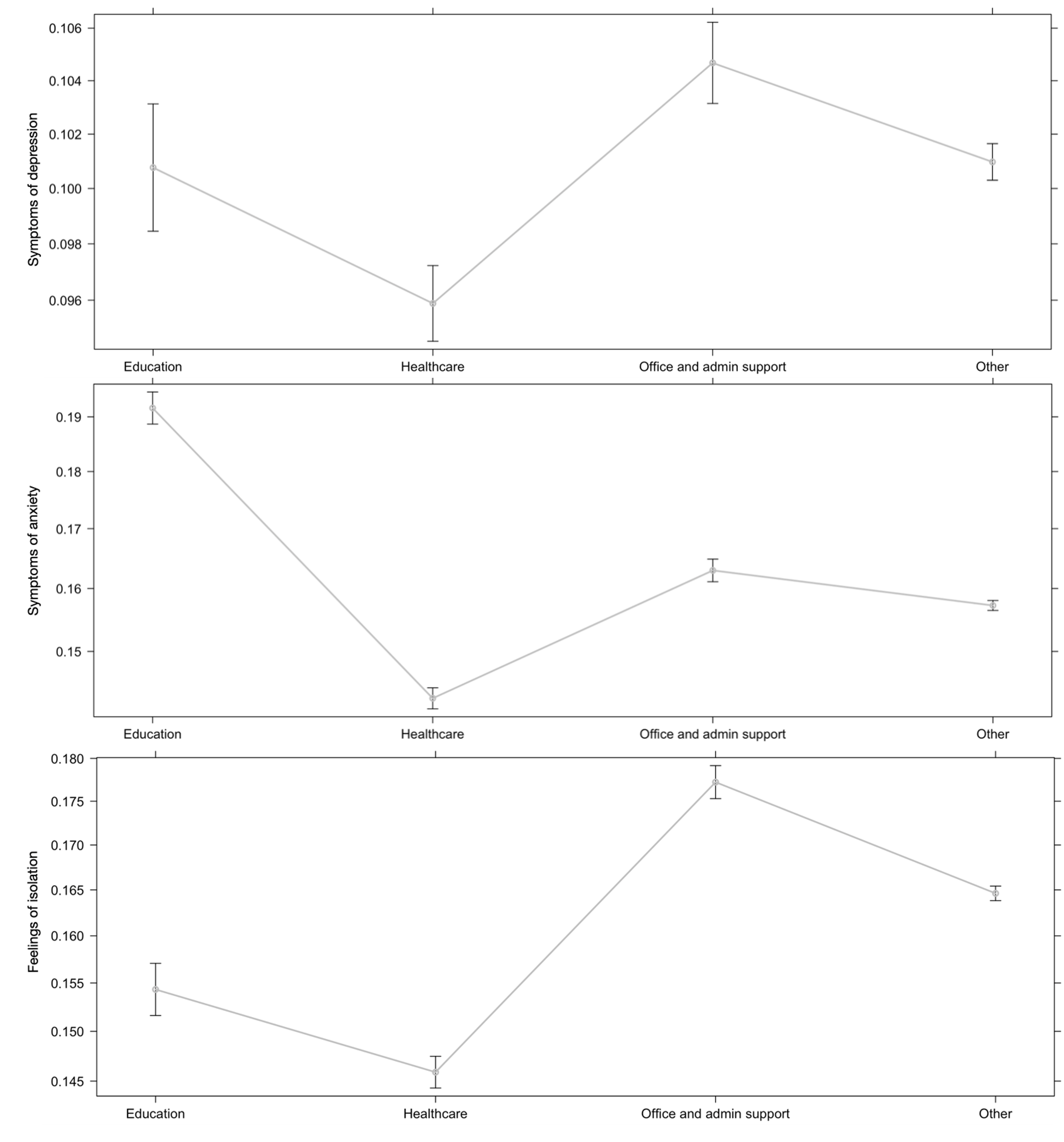}
  \begin{tablenotes}
  \small
  \item \textit{Figure 1}. Model Adjusted Probability of Mental Distress: Teachers vs. Other Professionals
\end{tablenotes}
\end{figure}

%
\begin{table}[!ht]
\centering
\small
\resizebox{\textwidth}{!}{\begin{tabular}{llllllll}
\multicolumn{8}{l}{Table 3}                                                                                                                             \\
\multicolumn{8}{l}{Results for   Prevalence of Mental Health Outcomes (Logistic regression): In-person vs.   Remote Teaching Modality} \\
\hline \\
\multicolumn{2}{l}{Variable} &
  \multicolumn{2}{l}{Depressive Symptoms} &
  \multicolumn{2}{l}{Anxiety Symptoms} &
  \multicolumn{2}{l}{Feelings of Isolation} \\
\hline \\
\multicolumn{2}{l}{}                                                      & OR   & CI (95\%)        & OR   & CI (95\%)        & OR   & CI (95\%)        \\
\cline{3-8} \\
\multicolumn{2}{l}{Type of Teaching (reference:   In-person)}             &      &                  &      &                  &      &                  \\
                 & Remote                                                 & 0.92 & {[}0.87, 0.98{]} & 1.01 & {[}0.96, 1.06{]} & 0.67 & {[}0.64, 0.70{]} \\
\multicolumn{2}{l}{Gender (reference:   Female)}                          &      &                  &      &                  &      &                  \\
                 & Male                                                   & 0.81 & {[}0.76, 0.88{]} & 0.56 & {[}0.53, 0.60{]} & 0.94 & {[}0.89, 1.00{]} \\
 &
  Non-binary &
  2.55 &
  {[}1.92, 3.39{]} &
  1.75 &
  {[}1.34, 2.27{]} &
  2.15 &
  {[}1.65, 2.80{]} \\
 &
  Other/prefer not to   answer &
  1.24 &
  {[}0.89, 1.72{]} &
  1.04 &
  {[}0.81, 1.34{]} &
  1.30 &
  {[}1.02, 1.67{]} \\
\multicolumn{2}{l}{Age (reference: 65+)}                                  &      &                  &      &                  &      &                  \\
                 & 18-24                                                  & 5.78 & {[}4.80, 6.96{]} & 7.01 & {[}6.03, 8.16{]} & 4.89 & {[}4.20, 5.71{]} \\
                 & 25-34                                                  & 3.87 & {[}3.30, 4.54{]} & 5.41 & {[}4.76, 6.15{]} & 3.32 & {[}2.93, 3.75{]} \\
                 & 35-44                                                  & 2.78 & {[}2.36, 3.26{]} & 3.82 & {[}3.36, 4.35{]} & 2.57 & {[}2.27, 2.90{]} \\
                 & 45-54                                                  & 2.21 & {[}1.88, 2.60{]} & 2.78 & {[}2.44, 3.16{]} & 2.03 & {[}1.79, 2.30{]} \\
                 & 55-64                                                  & 1.71 & {[}1.45, 2.01{]} & 1.91 & {[}1.67, 2.17{]} & 1.78 & {[}1.57, 2.02{]} \\
\multicolumn{2}{l}{Education Level (reference:   Less than HS)}           &      &                  &      &                  &      &                  \\
                 & HS                                                     & 0.28 & {[}0.13, 0.61{]} & 0.60 & {[}0.25, 1.39{]} & 0.31 & {[}0.14, 0.69{]} \\
                 & Some college                                           & 0.32 & {[}0.15, 0.69{]} & 0.71 & {[}0.31, 1.63{]} & 0.41 & {[}0.19, 0.88{]} \\
                 & College/professional   degree                          & 0.27 & {[}0.13, 0.59{]} & 0.80 & {[}0.35, 1.84{]} & 0.41 & {[}0.19, 0.90{]} \\
                 & Graduate degree                                        & 0.25 & {[}0.13, 0.59{]} & 0.76 & {[}0.33, 1.74{]} & 0.41 & {[}0.19, 0.89{]} \\
\multicolumn{2}{l}{Metro Size (reference:   Metro \textgreater 1 million} &      &                  &      &                  &      &                  \\
                 & Nonmetro-Not adjacent   to metro area                  & 0.89 & {[}0.79, 1.01{]} & 0.81 & {[}0.74, 0.89{]} & 0.85 & {[}0.77, 0.95{]} \\
                 & Nonmetro-Adjacent to   metro area                      & 0.91 & {[}0.83, 1.00{]} & 0.86 & {[}0.80, 0.92{]} & 0.92 & {[}0.85, 0.99{]} \\
                 & Metro-Fewer than   250,000                             & 1.00 & {[}0.91, 1.09{]} & 0.95 & {[}0.89, 1.02{]} & 0.95 & {[}0.88, 1.03{]} \\
                 & Metro-250,000 to 1   million population                & 1.02 & {[}0.96, 1.10{]} & 0.99 & {[}0.94, 1.04{]} & 1.04 & {[}0.98, 1.10{]}
 \\ 
\hline
\end{tabular}}

\begin{tablenotes}
  \small
  \item Note: OR = odds ratio. Only estimated coefficients for exposure and sociodemographic covariates are show. Models also adjusted for number of children and elders in the household, financial worry, and the number of COVID-19 positive people respondents knew, urbanicity, COVID-19 cases and deaths (lagged by 2 weeks), and fixed-effects for U.S. State and month.
\end{tablenotes}

\end{table}
%

\begin{figure}[!ht]
  \centering
  \includegraphics[width=1\columnwidth]{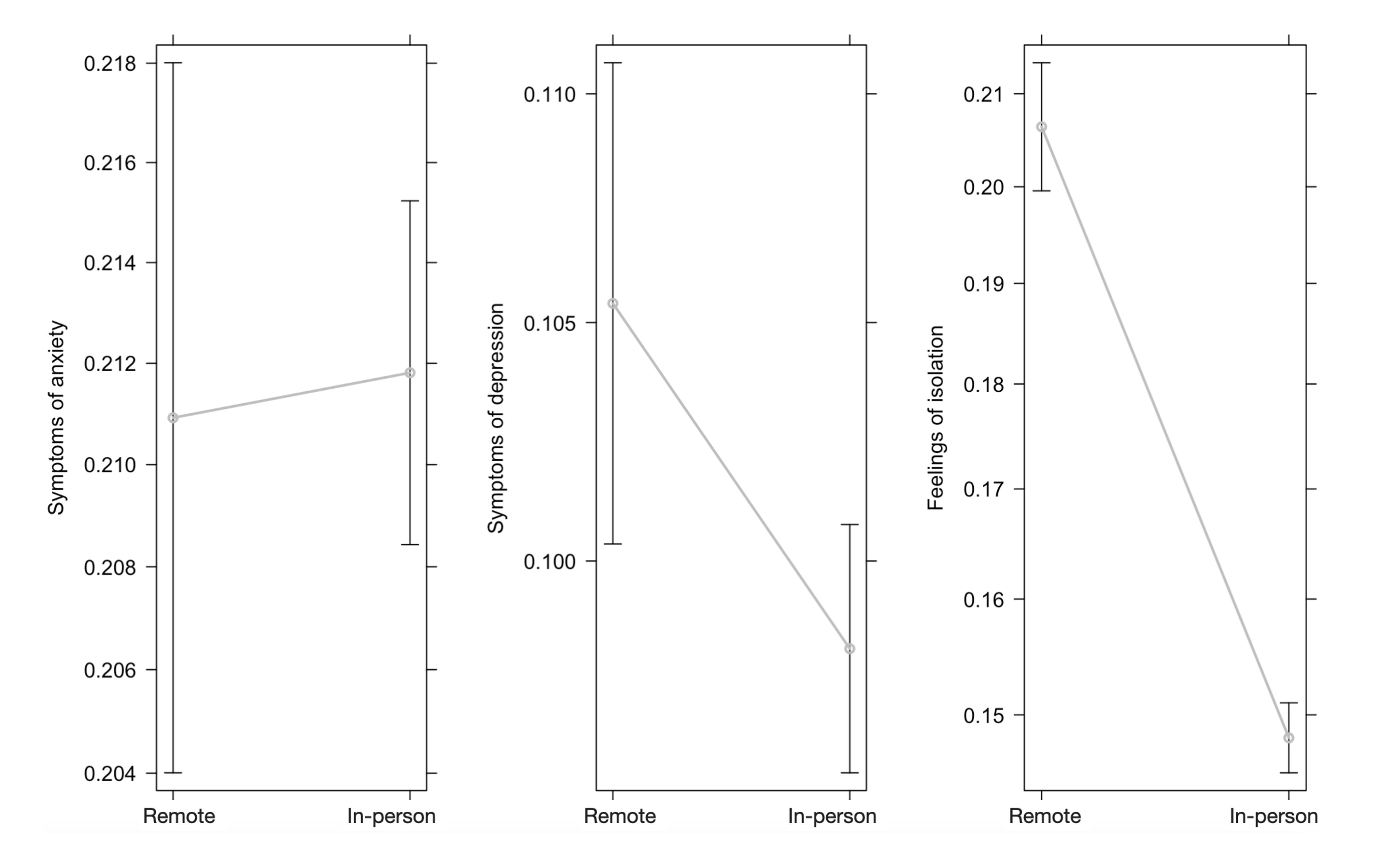}
  \begin{tablenotes}
  \small
  \item \textit{Figure 2}. Model Adjusted Probability of Mental Distress by Teaching Modality (In-person vs. Remote)
\end{tablenotes}
\end{figure}

\subsection{Comparison of Outcomes among In-Person and Remote Teachers}
\par
Among teachers, those teaching in-person were significantly less likely to report depressive symptoms (OR = 0.92, \textit{p} $<$ .001), and feelings of isolation (OR = 0.67, \textit{p} $<$ .001) than those teaching remotely. See Table 3 for full regression results and Figure 2 for a depiction of the adjusted probability of mental distress comparing those teaching in the different modalities. 

\section{Discussion}
\par
Teachers have faced unique challenges related to school reopening decisions due to the COVID-19 pandemic, such as high uncertainty on teaching modality changes, the need to abruptly adapt their lesson plans and teaching methods to remote-learning environments, and rapidly having to adopt new technologies to do so. In addition, teachers have seen their students struggle academically, socially, and/or emotionally with the added frustration of lacking the ability to provide them adequate support due to the pandemic-related disruptions (\shortcites{ferdig2020teaching}\citeauthor{ferdig2020teaching}, \citeyear{ferdig2020teaching}). Various guidelines have been proposed by organizations such as the \cite{centers2021guidance} and the \cite{national2020reopening} for safe and supportive learning environments as schools reopen, with reports often focusing on instructional modes of teaching or meeting students’ social-emotional needs. While teachers play a crucial role in such efforts, empirical studies assessing teacher mental health throughout the pandemic have been scarce. Our study addresses this gap, and indicates teachers have shown a significantly higher prevalence of negative mental health outcomes during the pandemic, compared to other professionals. Further, it shows those teaching remotely report significantly higher levels of distress than those teaching in person for all three mental health items considered in the study, even when controlling for individual sociodemographic variables and county-level COVID-19 spread.
This study is, to the best of our knowledge, the first to empirically assess these specific associations using a large national dataset. However, it is not without its limitations. Notably, the cross-sectional nature of the data and the limited covariates available precludes drawing causal inferences. Additionally, as baseline measures of pre-pandemic mental health outcomes are not available, it is not possible to compare the relationships found here with those pre-pandemic to determine if these findings are pandemic-specific.
More high-quality data and analyses are needed in order to assess the extent to which such heightened mental health distress, as well as its disparity between in-person and remote teachers, might be long-lasting. However, the connection between teacher and student emotional wellbeing – even during the COVID-19 pandemic – is well documented.

\subsection{Policy Implications}
\par
Although various guidelines have been proposed for safe and supportive learning environments as schools reopen (\citeauthor{EdResearch}, \citeyear{EdResearch}; \citeauthor{national2020reopening}, \citeyear{national2020reopening}; \citeauthor{centers2021guidance}, \citeyear{centers2021guidance}), these reports often failed to consider the magnitude and scope of possible negative effects on mental health outcomes among teachers, nor propose appropriate alternative methods and interventions to address such troubles. Following \shortcites{rossi2018evaluation}\cite{rossi2018evaluation}, we argue that incorporating information gathered from multiple stakeholders (including teachers) into decision-making processes is paramount for effective learning environments. This may be beneficial for two reasons in particular. First, \shortcites{harris2019teacher}\cite{harris2019teacher} have demonstrated that perceptions about work conditions for teachers (including teacher expectations, personal life issues, and job satisfaction) differed among principals, teachers, and other stakeholders. As a result, relying solely on input from principals or district leaders in decision-making processes may exclude the important, distinct perspectives of teachers. Second, the development of interpersonal relationships among teachers and other stakeholder groups may help reduce occupational stress among teachers \citep{segumpan2006teachers}. This relationship may be particularly important during the pandemic, given the likely increase in stress and negative mental health in teachers. 

\par 
There is strong evidence favoring public health containment measures such as partial or full school closures, in terms of their ability to help contain the spread of COVID-19. However, there are other societal factors and outcomes to consider along with their implementation. In this context, it is advisable to accompany these policies with interventions aiming to dampen some of their potential adverse effects. In particular, there is a need for tools and programs to support and safeguard the mental health of teachers both during and potentially after the pandemic. To do this effectively, the use of high-quality data collection is crucial. In this sense, our study focuses on teachers' mental health, leveraging a large national dataset to analyze outcomes for more than 130,000 teachers. Our findings, and this line of research more generally, have important potential implications for intervention research and practice to address teachers’ urgent mental health needs. Understanding the ways in which the mode of instruction during school reopening is related to teacher well-being can help inform interventions aimed to support and promote teachers’ mental health in this and future public health emergencies, ultimately supporting effective teaching practices. 


\newpage
\titleformat*{\section}{\bfseries\Large\centering} 
\setlength{\bibsep}{11pt} 
\bibliographystyle{apacite}
\bibliography{references2}
\end{document}